\documentclass{anstrans}
\title{Rossi-alpha Uncertainty Quanitification by Analytic, Bootstrap, and Sample Methods to Inform Fitting Best Practices}
\author{M.Y. Hua$^{1,2}$, C.A. Bravo$^{1,2}$, R.M. Marchie$^{1}$, J.D. Hutchinson$^2$, G.E. McKenzie$^2$, and S. A. Pozzi$^1$}

\institute{
1. Department of Nuclear Engineering and Radiological Sciences, University of Michigan, Ann Arbor, MI 48109\\
2. NEN-2: Advanced Nuclear Technology, Los Alamos National Laboratory, Los Alamos, NM 87545\\
}

\email{mikwa@umich.edu}

\usepackage{graphicx, color} % allows inclusion of graphics
\usepackage{booktabs} % nice rules (thick lines) for tables
\usepackage{microtype} % improves typography for PDF
\usepackage{units} %for \nicefrac
\usepackage{float}
\usepackage{enumerate}
\usepackage{amssymb, amsfonts, amsmath}
\usepackage{colortbl, multirow} 
\graphicspath{{Figures/}}

\allowdisplaybreaks

\begin{document}
%%%%%%%%%%%%%%%%%%%%%%%%%%%%%%%%%%%%%%%%%%%%%%%%%%%%%%%%%%%%%%%%%%%%%%%%%%%%%%%%
\section{Introduction and Motivation}
The prompt neutron period (the negative reciprocal of the prompt neutron decay constant) can be estimated using the Rossi-alpha technique that is predicated on fitting Rossi-alpha histograms and of interest in nuclear criticality safety and nonproliferation~\cite{Feynman44_1,Feynman44_2,Feynman56}.  The histograms are traditionally fit with a one-exponential model; however, recent work has proposed a two-exponential model to account for reflector-induced phenomenon \cite{Jesson2017,PHYSOR2006,mikwa_2exp}.  Until recently, the uncertainty quantification for either model was inadequate (inaccurate and demanded large measurement times).  Measurement uncertainty quantification by sample and analytic methods was developed and validated in Ref.~\cite{mikwa_UQ}.  The purpose of this transaction is to (i) validate a new bootstrap method by comparing bin-by-bin error bar estimates and (ii) demonstrate how to choose bin widths and reset times to optimize precision and accuracy.
 
%%%%%%%%%%%%%%%%%%%%%%%%%%%%%%%%%%%%%%%%%%%%%%%%%%%%%%%%%%%%%%%%%%%%%%%%%%%%%%%%
\section{Background}
Rossi-alpha histograms for bare-metal assemblies are typically fit with a one-exponential model, 
\begin{equation}
	p(t)dt = Adt + Be^{\alpha t}dt,
\end{equation} 
where the $A$ term represents uncorrelated counts that have a uniform probability of being detected at any time separation and the $B$ term represents correlated counts (same fission chain) that follow a decaying exponential trend \cite{Feynman44_1,Feynman44_2,Feynman56}.  The exponent, $\alpha$, is the prompt neutron decay constant.  Seminal work discusses further, observable correlations in Rossi-alpha histograms due to the presence of reflectors \cite{Avery,Cohn}, recent work shows that a two-exponential model better fits Rossi-alpha histograms of reflected assemblies \cite{Jesson2017,PHYSOR2006}, and prior work developed the associated two-exponential model from two-region point kinetics~\cite{mikwa_2exp}.  The two-exponential model,
\begin{equation}
	p(t)dt = Adt + B\left[\rho_1 e^{r_1t}+\rho_2 e^{r_2t}\right]dt 
\end{equation}
has two exponents and the prompt neutron decay constant is a linear combination of the two:
\begin{equation}
	\alpha = r_1(1-R) +r_2(R),
\end{equation}  
where $R$ is a determinable parameter between 0 and 1.  Further comprehensive details are given in Ref.~\cite{mikwa_2exp}.

The details of the uncertainty quantification by the analytic and sample methods is similarly left to citation in Ref.~\cite{mikwa_UQ}.  In essence, both methods aim to first estimate the uncertainty in the bin counts of the histogram of time differences.  The sample method splits a long measurement into several (at least 20) measurements, calculates histograms for the smaller measurements, then takes a sample standard deviation bin-by-bin.  The analytic method uses a fit to infer bin-specific Gaussian spreads and a binomial model to estimate bin-by-bin error bars (standard deviations).  The sample method is well defined and taken as the ground truth or reference.  Once the bin-by-bin error bars are obtained, the uncertainty is propagated through the fit algorithm by way of weighting.  In the case of the two-exponential model, the resulting uncertainty in fit parameters is propagated to the final estimate of the prompt neutron period.

%%%%%%%%%%%%%%%%%%%%%%%%%%%%%%%%%%%%%%%%%%%%%%%%%%%%%%%%%%%%%%%%%%%%%%%%%%%%%%%%
\section{Rossi-alpha Bootstrapping Algorithm}
A bootstrapping algorithm is based on sampling subsets of a total dataset; for Rossi-alpha measurements, we resample the sorted list of neutron detection times (list mode data). A set of 1,000 subsequent times is obtained by randomly selecting the first time and sets are collected and combined until the length of the list is equal to the original data set.  This final list constitutes a single sample and the size of the sets -- 1,000 -- is referred to as stride length.  Stride lengths should correspond to net time differences longer than typical fission chain lengths and may be affected by amounts of background radiation; it is preferential to choose larger times as opposed to smaller ones, though arbitrarily large stride lengths will detrimentally affect precision.  We resample the data 10,000 times in this work and a Rossi-alpha histogram is created for each resample.  The variance-covariance matrix (variance on the diagonal and covariance on the off-diagonal terms) for a given histogram (the counts in each bin, thus the diagonal of the variance-covariance matrix is the errorbar squared) is calculated from the 10,000 resamples.  The error bars are then used to weight the fitting algorithm as described in Ref.~\cite{mikwa_2exp}.  Note that calculating the prompt period for each resample and then taking a sample standard deviation is not an equivalent bootstrapping method since the unweighted fits are less accurate and do not adequately account for the histogram uncertainty.

%%%%%%%%%%%%%%%%%%%%%%%%%%%%%%%%%%%%%%%%%%%%%%%%%%%%%%%%%%%%%%%%%%%%%%%%%%%%%%%%
\section{Measurement}
\begin{figure}[H]
	\centering
	\includegraphics[width=\linewidth]{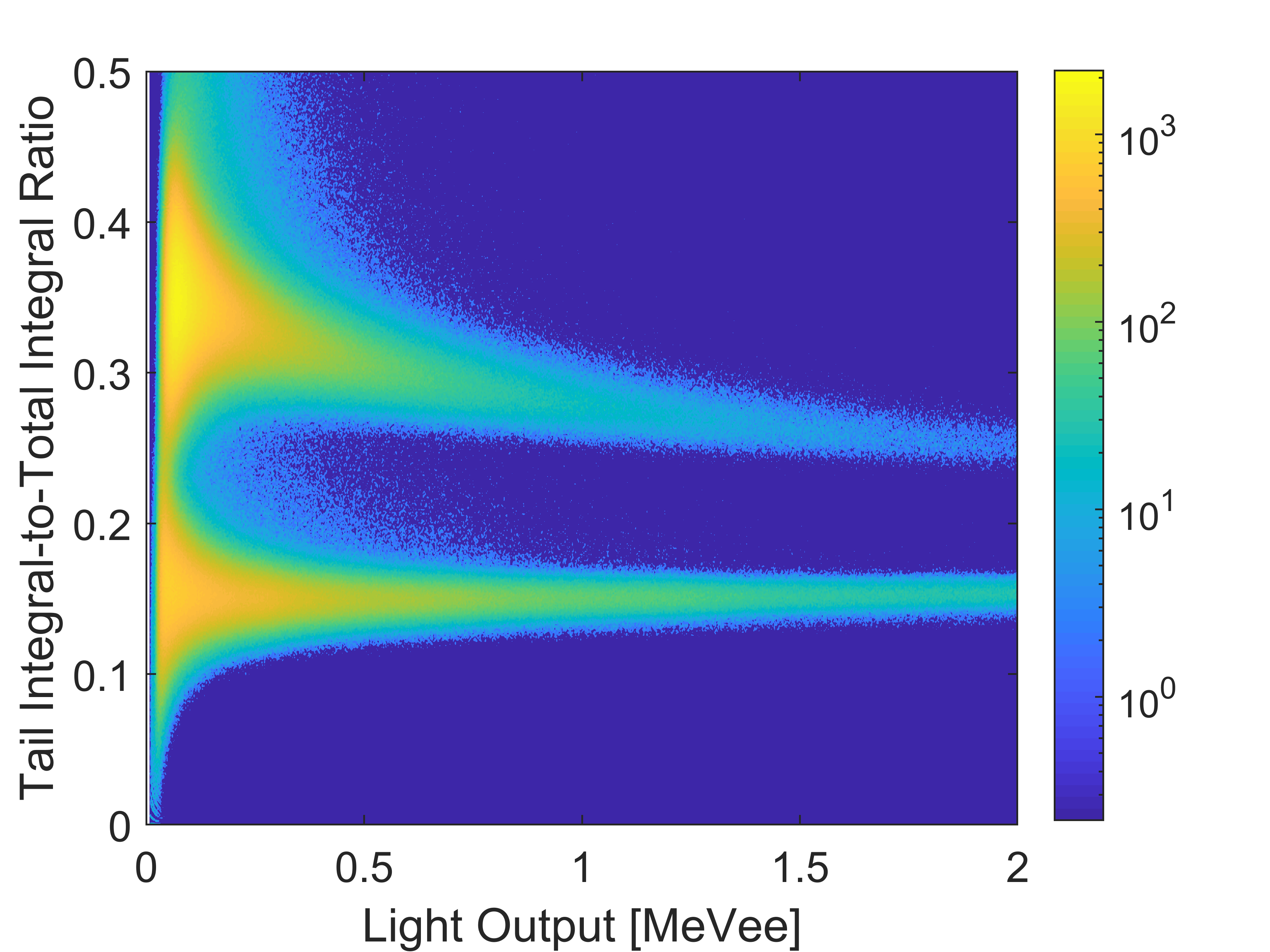}
	\caption{Sample pulse shape discrimination plot for the BeRP ball reflected by 10.16 cm of copper.}
	\label{fig:PSD}
\end{figure}

The measurement consisted of 12 \textit{trans}-stilbene organic scintillation detectors measuring a 4.5-kg sphere of weapons-grade, alpha-phase plutonium -- known as the BeRP ball -- reflected by 7.62 cm of nickel, copper, or tungsten, or 10.16 cm of copper.  The measurement is identical to that of Ref.~\cite{RA_w_Organics}, which validates the use of fast organic scintillators in Rossi-alpha measurements.  Organic scintillators are sensitive to neutrons and photons, and the pulses are distinguished using pulse shape discrimination.  A sample plot is shown in Fig.~\ref{fig:PSD}.  The output of preliminary data processing is list mode data, or a list of neutron detection times.  Comprehensive measurement details and preprocessing steps are discussed in Refs.~\cite{mikwa_UQ,RA_w_Organics}.

%%%%%%%%%%%%%%%%%%%%%%%%%%%%%%%%%%%%%%%%%%%%%%%%%%%%%%%%%%%%%%%%%%%%%%%%%%%%%%%%
\section{Results and Discussion}
\begin{figure}[H]
	\centering
	\includegraphics[width=\linewidth]{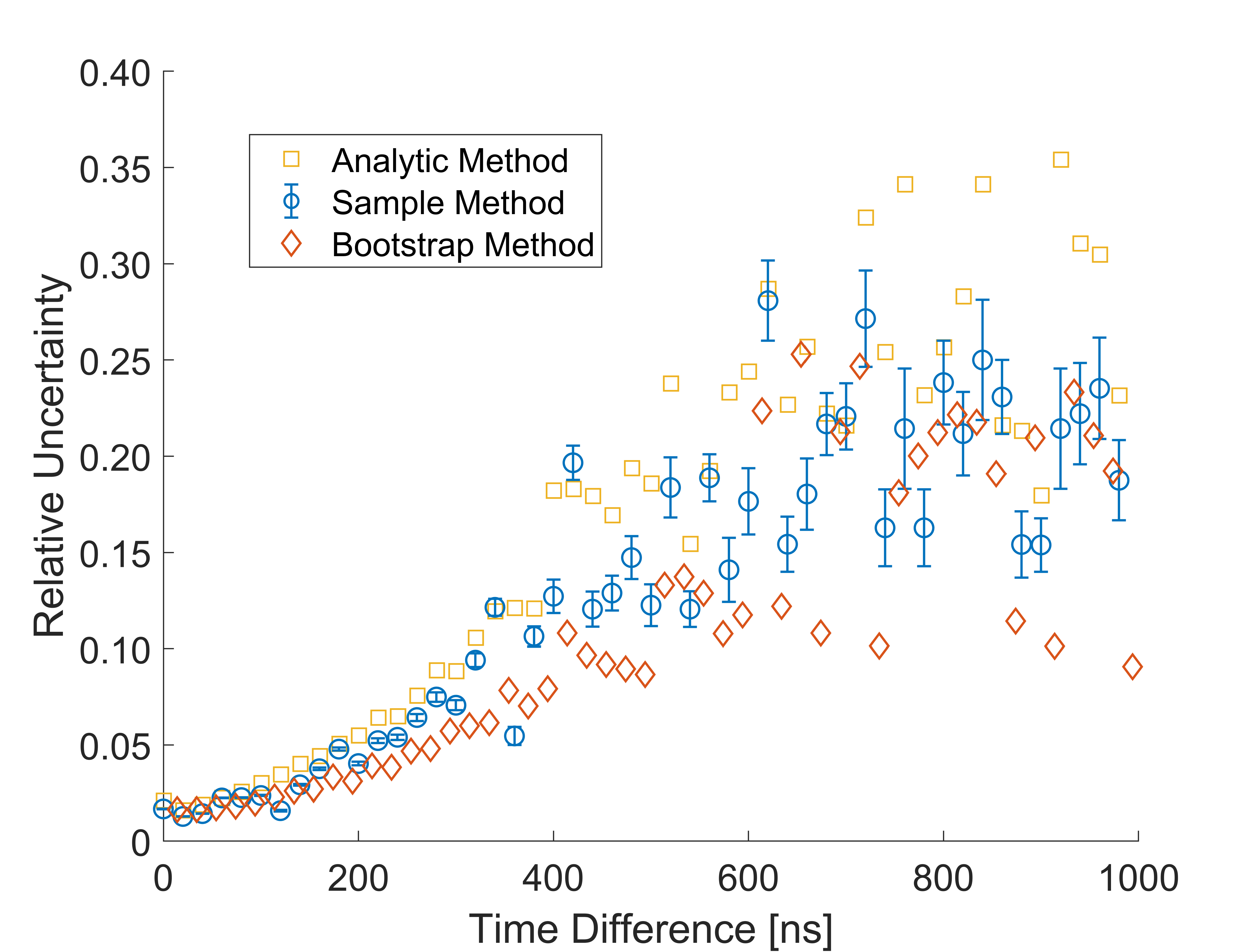}
	\caption{Validation of the analytic and bootstrap estimates of bin-by-bin uncertainty for the BeRP ball reflected by 7.62 cm of copper.  The x-axis are all neutron time differences less than 1000 ns between any and all neutron detections.  Only every 20 points are plotted for clarity.}
	\label{fig:direct}
\end{figure}

\begin{figure}[H]
	\centering
	\includegraphics[width=\linewidth]{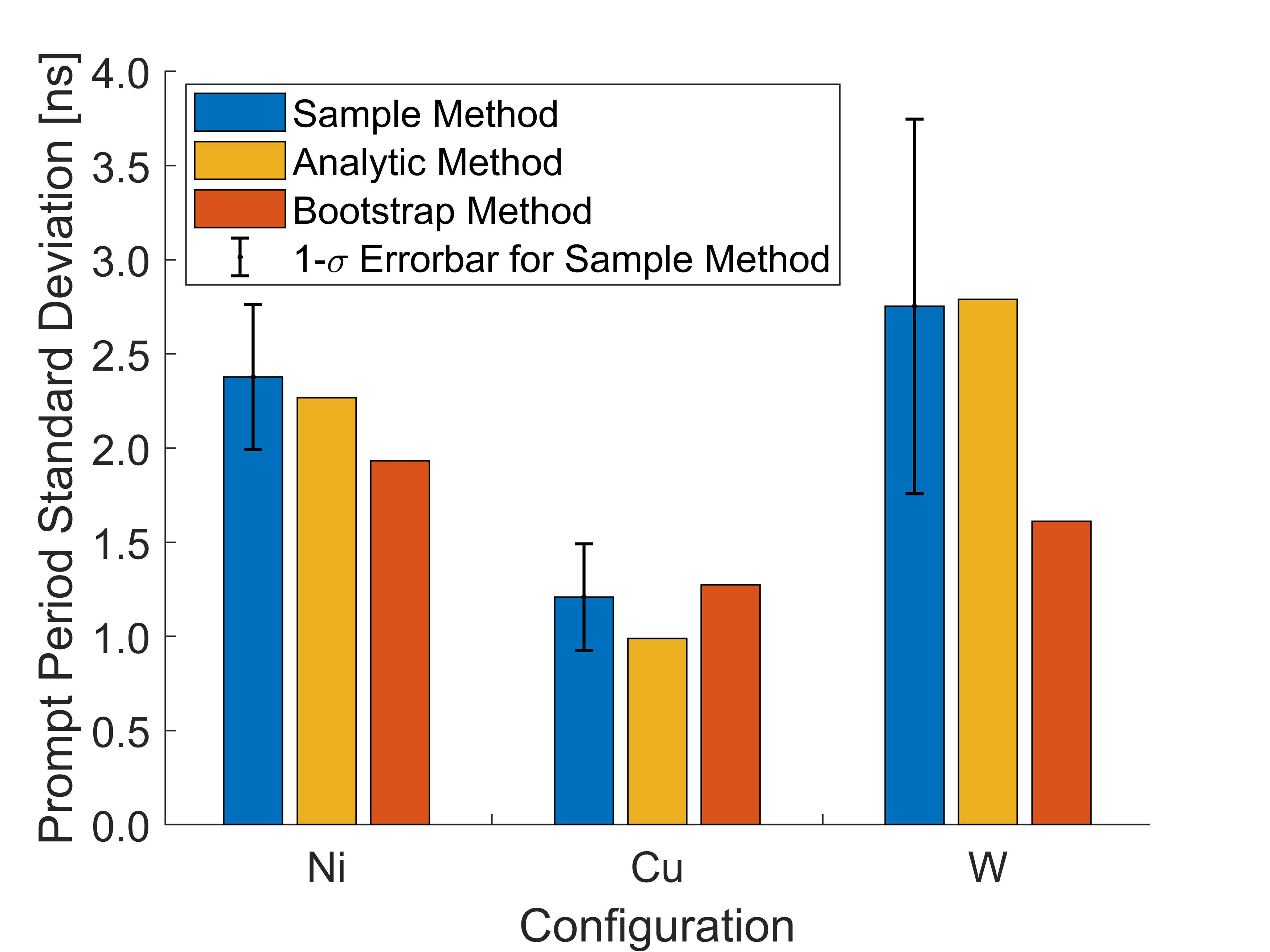}
	\caption{Validation of analytic and bootstrap methods for the BeRP ball and 7.62 cm reflectors.}
	\label{fig:validation}
\end{figure}

\begin{figure}[H]
	\centering
	\includegraphics[width=\linewidth]{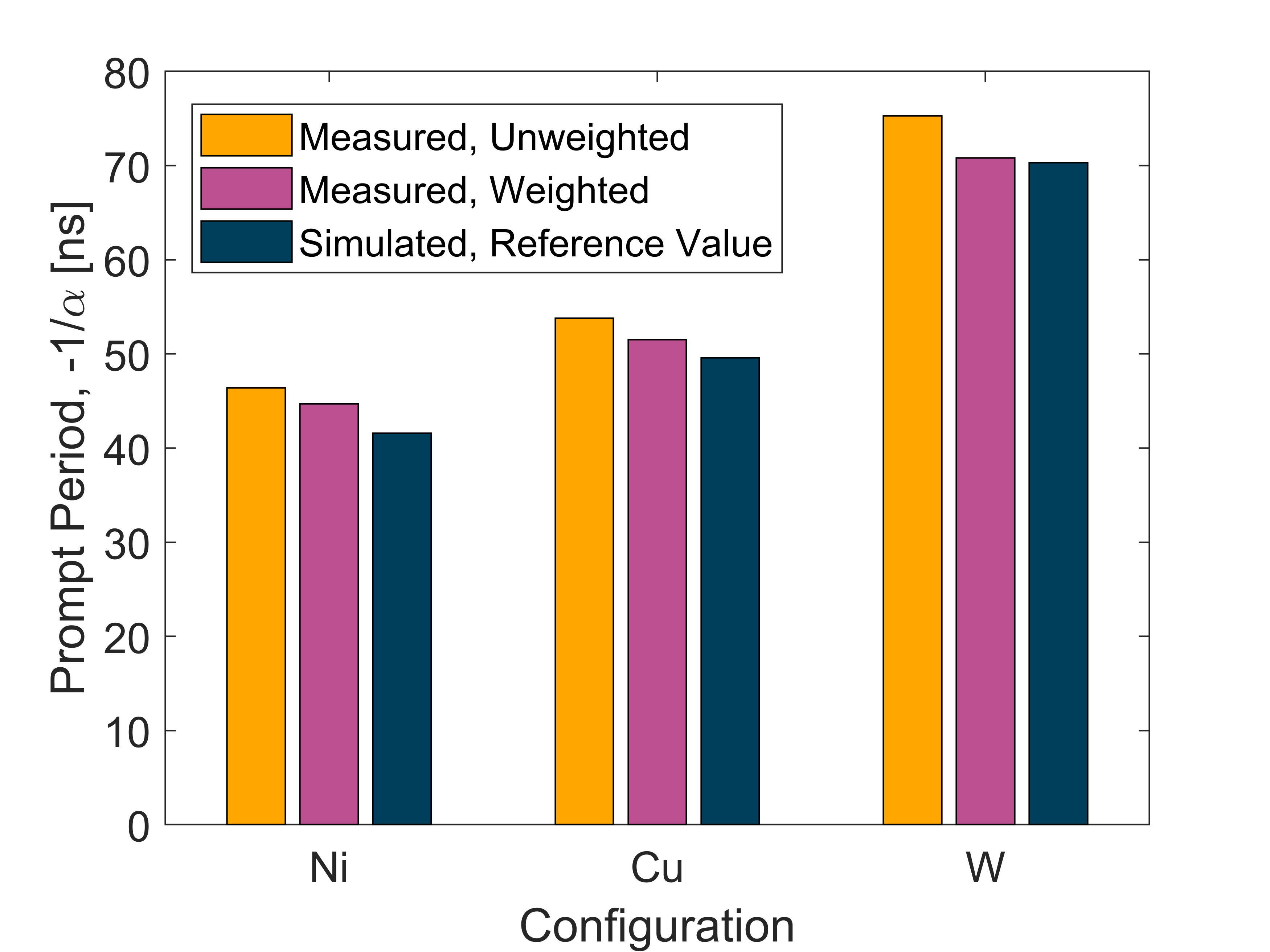}
	\caption{Demonstration of accuracy improvement due to weighting for the BeRP ball and 7.62 cm reflectors.  Figure from Ref.~\cite{mikwa_UQ}.}
	\label{fig:improvement}
\end{figure}

The (not-yet-optimized) bootstrap method is validated by comparing bin-by-bin relative uncertainty to the reference, ground-truth sample method, shown in Fig.~\ref{fig:direct}.  Note that the bootstrap method is less conservative than the analytic method.  The histogram uncertainty is propagated to the final estimate of the prompt neutron period and the resulting error bars are compared in Fig.~\ref{fig:validation}; the good agreement is expected based on the good agreement shown in Fig.~\ref{fig:direct}.  The improvement in accuracy is due to weighting the fits is shown in Fig.~\ref{fig:improvement}.

\begin{figure}[H]
	\centering
	\includegraphics[width=\linewidth]{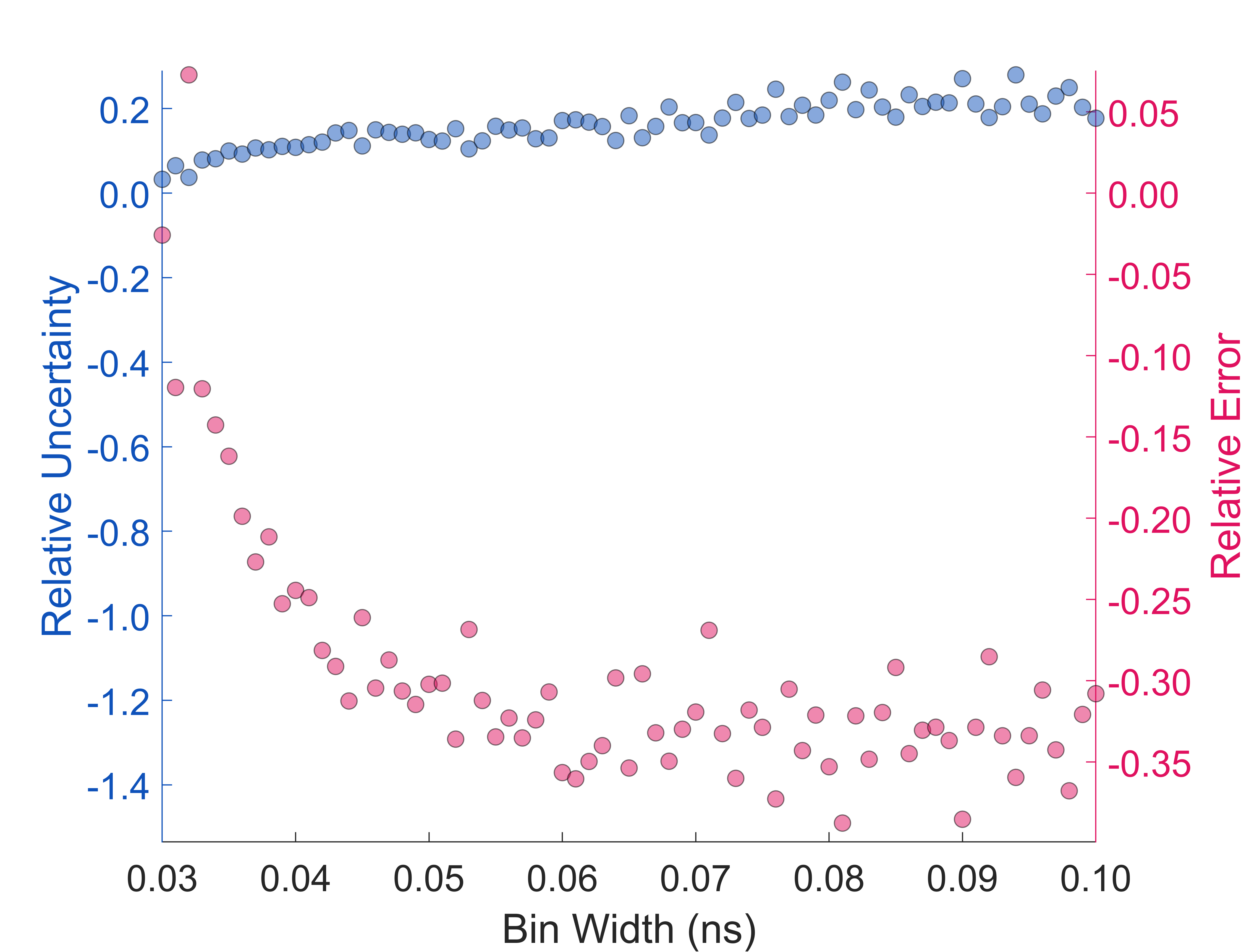}
	\caption{Sample bin-width optimization for the BeRP ball reflected by 10.16 cm of copper.}
	\label{fig:bins}
\end{figure}

\begin{figure}[H]
	\centering
	\includegraphics[width=\linewidth]{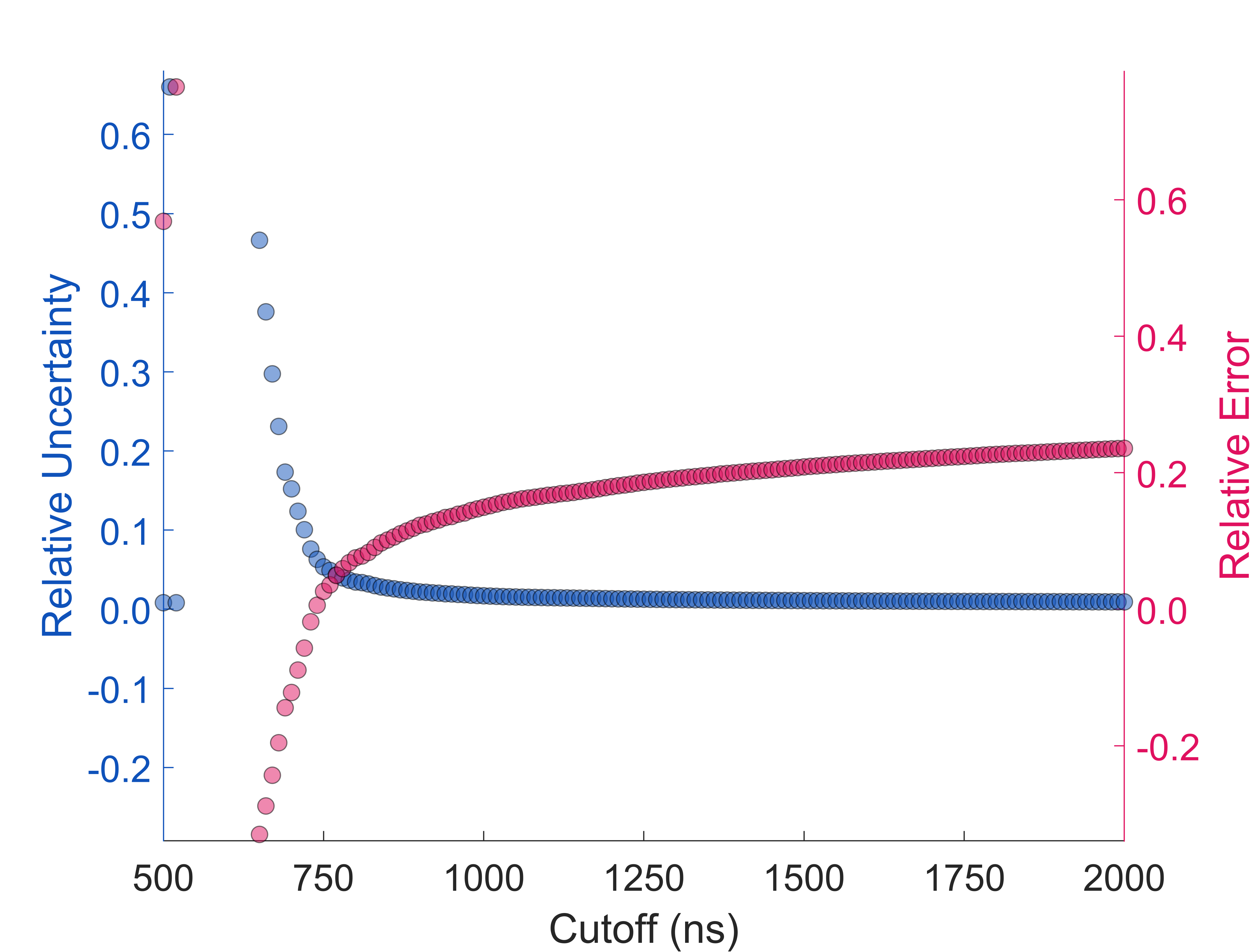}
	\caption{Sample tail-length optimization for the BeRP ball reflected by 7.62 cm of tungsten.}
	\label{fig:tail}
\end{figure}

\begin{figure}[H]
	\centering
	\includegraphics[width=\linewidth]{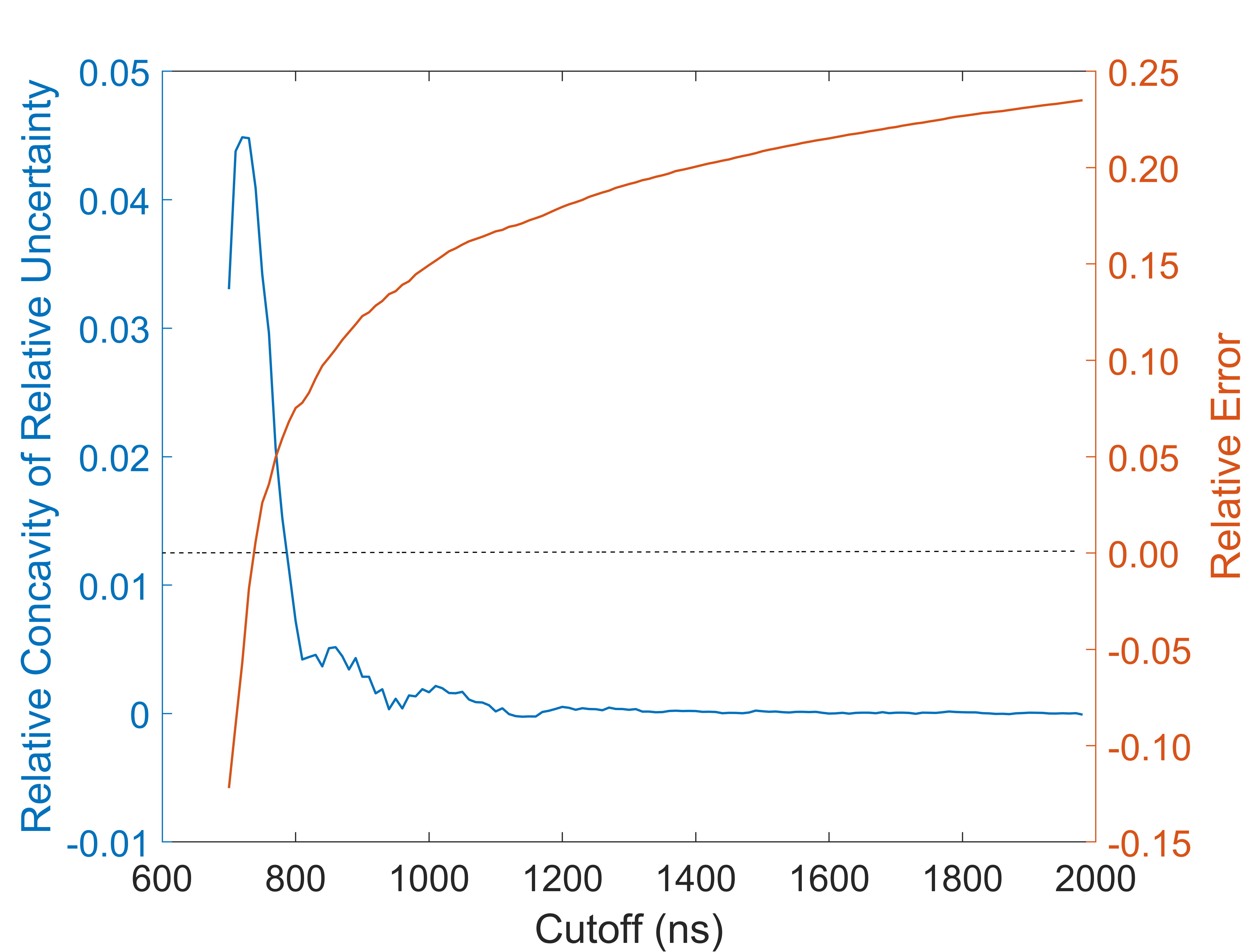}
	\caption{Sample relative curvature plot to determine optimal cutoff for the BeRP ball reflected by 7.62 cm of tungsten.}
	\label{fig:curvature}
\end{figure}

The uncertainty estimates are used to determine fitting best practices that optimize precision and accuracy.  Two common parameters include the histogram bin width and the reset time (maximum time difference to record).  The relative uncertainty and relative error as a function of bin width for a fixed 1000-ns reset time are shown in Fig.~\ref{fig:bins}.  The bin width is bounded below by the effective clock tick length of the electronics; in this work, the lower bound is 0.03 ns.  It has been observed that the minimum bin width typically results in the lowest relative error \textit{and} relative uncertainty.  When the minimum bin width is not optimal, a similar process to the optimization of reset time, shown below, should be performed.  Reducing bin size reduces precision in the count in each bin.  The improved merit of smaller bin sizes indicates that the reduction in uncertainty/error due to more points for the fitting algorithm to use is greater than the increase in uncertainty/error due to poorer statistics in each bin. 
The relative uncertainty and relative error as a function of reset time for a fixed 0.03-ns bin width are shown in Fig~\ref{fig:tail}.  The trend in relative uncertainty suggests that arbitrarily large reset times are preferential as the uncertainty tends to zero; however, the magnitude of the relative error is minimized at the elbow of the relative uncertainty curve.  Since the relative error is unknown in practice, the relationship between the two curves must be exploited.  The elbow of the relative uncertainty curve is determined by finding the index of the peak of the second derivative divided by the relative uncertainty, or the relative curvature.  The relative curvature of the relative uncertainty curve is overlayed with the relative error curve in Fig.~\ref{fig:curvature}, therein demonstrating the method.  Note that this method assumes that the relative uncertainty is sufficiently small.  If greater precision is desired, longer measurements should be acquired if possible since using larger reset times offers diminishing returns.
%%%%%%%%%%%%%%%%%%%%%%%%%%%%%%%%%%%%%%%%%%%%%%%%%%%%%%%%%%%%%%%%%%%%%%%%%%%%%%%%
\section{Conclusion, and Future Work}
The bootstrap method is validated by comparison to the sample method, though we note that it is not a conservative estimate like the analytic method is.  The error bars are propagated to the uncertainty in the prompt period, which are shown to agree with the reference, sample method values within 95.4\% confidence intervals.  Weights should be used when fitting Rossi-alpha histograms to properly propagate uncertainty and to improve accuracy.  Any of the three uncertainty methods, shown to give equivalent results, should be used to inform optimal bin widths and reset times, which depend on the measurement system and the assembly being measured.

Based on observation in the developmental phase of this work, the bootstrapping algorithm is moderately sensitive to stride length and reset time.  The 10,000 resamples was taken from reference and the 1,000 stride length was arbitrarily chosen to be much larger than typical fission chain lengths (a stride length of 20 is sufficient for our system and a moderated system would require many more).  A parametric study of the bootstrap method as a function of stride length and reset time are the subject of future work.  Similar to the Feynman-Y method that utilizes the deviation of data from a Poisson distribution, we are interested in investigating the feasibility of using the optimal bin width or reset time as a signature.  A numerical derivative is used in this work to determine the optimal reset time.  Two derivatives means that the index could be off by up to two units of $\Delta$x, which is 10 ns in this work.  In the future, we will fit the data and take a functional derivative.  If well-behaved fits are not available, we will use an Euler optimization scheme to zoom into optimal reset times until the electronic limit is reached.  Lastly, we intend to compare the analytic, sample, and bootstrap methods, particularly as a function of total measurement time.  Currently, it has been shown that the analytic and sample methods are more desirable for on-the-fly analysis due to computational time (though the bootstrap method could be improved with GPU and matrix programming). 

\section{Acknowledgments}
This work was partially supported by the  National Science Foundation Graduate Research Fellowship under Grant No. DGE-1256260, the Consortium for Verification Technology under Department of Energy National Nuclear Security Administration award number DE-NA0002534, the Consortium for Monitoring, Technology, and Verification under Department of Energy National Nuclear Security Administration award number DE-NA0003920, and the DOE Nuclear Criticality Safety Program, funded and managed by the National Nuclear Security Administration for the Department of Energy.  Any opinion, findings, and conclusion or recommendations expressed in this material are those of the authors and do not necessarily reflect the views of any funding organization.  

%%%%%%%%%%%%%%%%%%%%%%%%%%%%%%%%%%%%%%%%%%%%%%%%%%%%%%%%%%%%%%%%%%%%%%%%%%%%%%%%
\bibliography{bibliography}
\bibliographystyle{ans}
\end{document}